\documentclass[prl,twocolumn,letter]{revtex4-1}
\usepackage{bm,graphicx,graphics,amsmath,amssymb,bm,epsfig,color}
\usepackage{euscript,tabularx}
\usepackage{longtable}
\begin{document}

\title{Reply to the comment}

\author{C. Hahn}
\author{G. de Loubens}
\author{M. Viret}
\author{O. Klein}\email[Corresponding author:]{ oklein@cea.fr}

\affiliation{Service de Physique de l'\'Etat Condens\'e (CNRS URA
  2464), CEA Saclay, 91191 Gif-sur-Yvette, France}

\author{V.V. Naletov}

\affiliation{Service de Physique de l'\'Etat Condens\'e (CNRS URA
  2464), CEA Saclay, 91191 Gif-sur-Yvette, France \\ Institute of
  Physics, Kazan Federal University, Kazan 420008, Russian Federation}

\author{J. Ben Youssef}

\affiliation{Universit\'e de Bretagne Occidentale, Laboratoire de
  Magn\'etisme de Bretagne CNRS, 6 Avenue Le Gorgeu, 29285 Brest,
  France}

\maketitle

%pacs 85.75.-d, 76.50.+g

The comment of M. Weiler \textit{et al.} \cite{Weiler2014a} raises a
pertinent question about the origin of the electrical signal detected
in our letter \cite{Hahn2013}, reporting on the detection of the ac
part of the spin-pumping current emitted during ferromagnetic
resonance using the inverse spin Hall effect (ac-ISHE). The
originality of our method was to induce a resonance in YIG{\textbar}Pt
at half the frequency using parametric excitation in the parallel
geometry. Other attempts to measure the ac-ISHE have used a balanced
circuit \cite{Wei2013}, spin rectification effects \cite{Hyde2013} or
phase detection \cite{Weiler2014}. M. Weiler \textit{et al.} point out
to an inconsistency in the interpretation of our data: if indeed the
uniform mode of our YIG would be excited, then the produced
ferromagnetic inductive (FMI) voltage should have dominated over the
ac-ISHE voltage and it should have lead to the same signal amplitude
in both YIG{\textbar}Pt and YIG{\textbar}Al. In ref. \cite{Hahn2013},
we report a signal in YIG{\textbar}Pt that is about an order of
magnitude smaller than the predicted FMI-voltage and the signal
vanishes in the case of YIG{\textbar}Al, where only the FMI-voltage
should dominate.

Because we missed the estimation of the expected FMI contribution, we
have revisited exhaustively \footnote{We have checked that the
  experimental results presented herein are similar on different
  geometries, using different NM patterns (single slab or double
  coplanar loop), flipping sample side, detecting the signal at $f/2$
  or $3f/2$, as well as changing the YIG thickness.}  our measurement
of the ratio, $\rho$, between the signal measured in the
YIG{\textbar}Pt and YIG{\textbar}NM, where NM is a normal metal
suppressing the ISHE. We present in FIG.1 two sets of measurements
performed near the onset of the parametric excitation. We first show
in panel (b) and (c), a comparison of the signal measured in
YIG{\textbar}Pt$_\text{7nm}$ and in
YIG{\textbar}Pt$_\text{7nm}${\textbar}Al$_\text{50nm}$
\footnote{Measurement of the dc-ISHE confirms that the addition of the
  50nm Al layer is sufficient to shortcut spin-orbit effect in the
  NM.}. The two sets are performed at the same YIG location, ensuring
that the parametric threshold is unchanged. Although increasing the NM
thickness reduces the impedance of the circuit, this should enhance
the FMI-part of the signal. This method thus yields an
under-estimation of the ratio $\rho= {\left | V_{\rm iSHE} + V_{\rm
      FMI} \right |}/{\left | V_{\rm FMI} \right |} $. Comparing the
two samples, we find that the ratio $\rho$ is larger than 5 when $P\le
24$~dBm. The disappearance of the signal above 3~GHz is due to the
fact that the stripline becomes there inefficient to pump
parametrically the YIG. In order to check the influence of the
impedance match, we have repeated the measurement in another YIG
sample covered by three electronically connected slabs of respectively
Pt$_\text{7nm}$, Al$_\text{15nm}$ and Pt$_\text{7nm}$. Displacing the
antenna laterally above each slab allows to selectively excite the 3
different regions using the same impedance circuit (see FIG1
(d-f)). Although spatial variation of the YIG quality leads to error
bars in the estimation of $\rho$, the 3 sets of measurements confirm
that $\rho>1$ at the settings used in ref. \cite{Hahn2013}. We
also observe that this result is specific to excitation near the
parametric threshold: at much larger power the ratio $\rho$ decreases
and eventually reaches 1.

Although, the large measured value of $\rho$ confirms
\emph{experimentally} that indeed the ac-ISHE has been detected in
ref. \cite{Hahn2013}, we need to reconcile this result with the
prediction \cite{Weiler2014a} of the dominance of the FMI-voltage
($\rho=1$). We note that $\rho -1 \propto {\langle \dot{M}_y
  \rangle}/{ \langle \dot{M}_x \rangle} $, where the over-dot denotes
the time derivative and the chevron bracket indicates the spatial
average (see FIG.1(a) for axes orientation). Therefore standing
spin-waves (SW) excited along the YIG film thickness direction can
significantly decrease the value of the FMI-voltage, while leaving the
ac-ISHE voltage unchanged. Moreover, parametric excitation is known to
be efficient at exciting spatially inhomogeneous SW \cite{sparks64,
  Vendik1977, Guo2014}. Usually, the first magnons to go unstable are
the $\pi/2$-magnons ($k\perp M$) \cite{sparks64}. While we could
determine experimentally that the excited SW are inhomogeneous in the
$x-z$ plane \footnote{patterning the NM into a coplanar double loop of
  size $w$, increases the FMI-sensitivity to $k\sim 1/w$.}, we
could not find an unambiguous way to demonstrate that they are also
inhomogeneous along $y$ in our 200~nm thick YIG. This thus leaves us
with a plausible explanation of the discrepancy but without a direct
proof that this is what indeed occurs.

In summary, we confirm that close to the threshold the ratio $\rho$ is
larger than 1, suggesting that the signal is indeed dominated by the
ac-ISHE as reported in our paper \cite{Hahn2013}. Nevertheless, the
fact that uncharacterized spatially inhomogeneous SW are excited
prevents a quantitative analysis of the measured signal.

\begin{figure}
  \includegraphics[width=8.5cm]{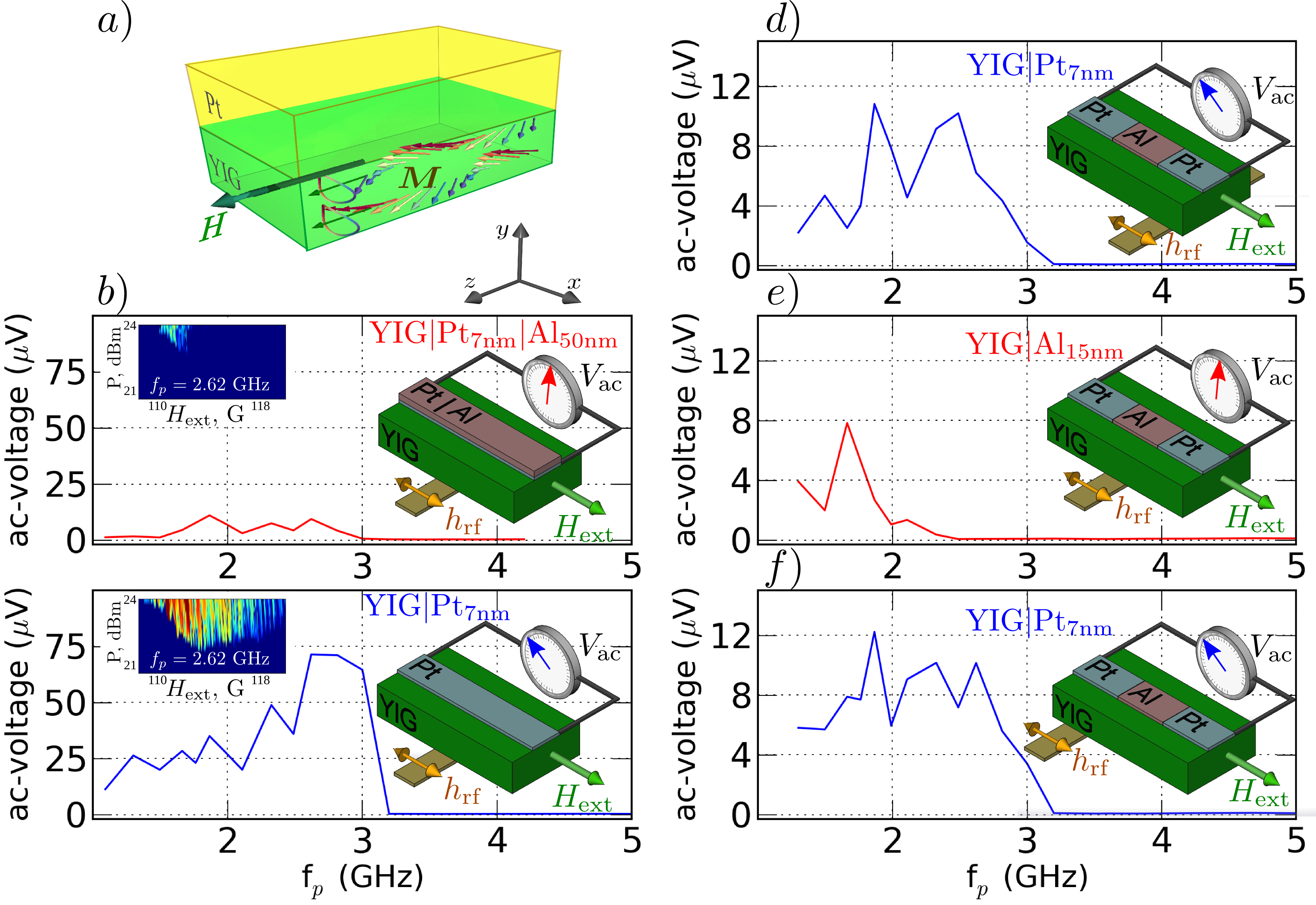}
  \caption{(Color online) (a) Measurement of the ac-voltage
    ($V_\text{ac}$) produced by spin-waves excited parametrically. (b)
    and (c) are a comparative study of
    YIG{\textbar}Pt$_\text{7nm}${\textbar}Al$_\text{50nm}$ and
    YIG{\textbar}Pt$_\text{7nm}$ where the same YIG location is
    excited.  $V_\text{ac}$ as function of power ($P$) and bias
    magnetic field ($H_\text{ext}$), at constant pumping frequency
    ($f_p$), is shown in the insert. From there, the maximum
    $V_\text{ac}$ for $P\le 24$~dBm is extracted and plotted as
    function of $f_p$. Same measurement done on a slab covered
    successively by Pt$_\text{7nm}$(d) / Al$_\text{15nm}$(e) /
    Pt$_\text{7nm}$(f). Displacing the microwave antenna underneath
    the YIG allows to excite successively the Pt and Al using the same
    impedance circuit.}
  \label{FIG1}
\end{figure}

%\bibliography{YIG}
%Merlin.mbs v4.21 2009-07-09.
%

\end{document}